Low-cost sensor networks and land-use regression: interpolating nitrogen dioxide concentration at high temporal and spatial resolution in Southern California.

Lena Weissert[1,3,a], Kyle Alberti[2], Elaine Miles[2], Georgia Miskell[1b], , Brandon Feenstra[4], Geoff S Henshaw[2], Vasileios Papapostolou[4], Hamesh Patel[2], Andrea Polidori[4], Jennifer A Salmond[3], David E Williams[1,*].

*Email  david.williams@auckland.ac.nz    ph +64 9 923 9877

1. School of Chemical Sciences and MacDiarmid Institute for Advanced Materials and Nanotechnology, University of Auckland, Private Bag 92019, Auckland 1142, New Zealand

2. Aeroqual Ltd, 460 Rosebank Road, Avondale, Auckland 1026, New Zealand

3. School of Environment, University of Auckland, Private Bag 92019, Auckland 1142, New Zealand

4. South Coast Air Quality Management District, 21865 Copley Drive, Diamond Bar, CA 91765, USA

*Abstract*

The development of low-cost sensors and novel calibration algorithms offer new opportunities to supplement existing regulatory networks to measure air pollutants at a high spatial resolution and at hourly and sub-hourly timescales. We use a random forest model on data from a network of low-cost sensors to describe the effect of land use features on local-scale air quality, extend this model to describe the hourly-scale variation of air quality at high spatial resolution, and show that deviations from the model can be used to identify particular conditions and locations

[a] Present address:  Aeroqual Ltd, 460 Rosebank Road, Avondale, Auckland 1026, New Zealand
[b] Present address: Trustpower, 108 Durham St, Tauranga, New Zealand

where air quality differs from the expected land-use effect. The conditions and locations under which deviations were detected conform to expectations based on general experience.

*Introduction*

The South Coast Air Basin is one of the most polluted air basins in the United States (Epstein et al., 2017). The pollution problem in this region is driven by high emissions, unfavourable meteorological conditions (low wind speed, strong temperature inversions, abundant sunshine, infrequent rainfall), sea breezes and complex terrain that limits pollutant dispersion (South Coast AQMD, 2016). Spatially and temporally dense information about local scale air pollution is necessary to mitigate air pollution effectively (Vizcaino and Lavalle, 2018). While regulatory air quality monitoring networks offer important insights about long-term air quality trends, the data must be supplemented with additional measurements and models to obtain geographically more detailed air pollution information (Li et al., 2019a). This is of importance given that air pollutants vary considerably over small distances (Kumar et al., 2015; Weissert et al., 2019a).

Association of average pollutant concentration with land use variables (e.g. distance to major roads, length of major roads within different buffers, bus stops) is a frequently used approach to model time-averaged pollutant concentrations with high spatial resolution (Hoek et al., 2008). A limitation of land use regression (LUR) models is the risk of overfitting the data when only few measurement sites are used to train the model. Further, LUR modelling is based on the assumption that relationships between air pollution and predictor variables are linear and that there are no interaction effects between different predictors. Given these limitations, researchers have considered using other algorithms that account for some of these limitations (e.g. Generalized Additive Model (GAM), Least Absolute Shrinkage and Selection Operator (LASSO)) to fit a land use model to pollutant concentrations (Chen et al., 2019). Some studies have also used machine learning algorithms such as Random Forest (RF) (Brokamp et al., 2017; Hu et al., 2017; Zhan et al., 2018). LUR models are usually developed from dense diffusion tube monitoring over a few weeks during different seasons and lack temporal resolution at the hourly or sub-hourly scale. To overcome this, LUR models have been combined with temporally variable predictors to obtain hourly models (Masiol et al., 2018; Miskell et al., 2018b; Son et al., 2018; Yeganeh et al., 2018).

The development of low-cost sensors has created new opportunities for air quality measurements and modelling. If deployed in dense networks, low-cost sensors have the potential to provide near real-time measurements of pollutants at a spatial resolution representative of the neighbourhood scale. They can offer insights into the influence of local pollution sources at different temporal and spatial scales that may not be detected by the usually sparsely distributed regulatory monitoring networks (Feinberg et al., 2019; Li et al., 2019b; Popoola et al., 2018; Weissert et al., 2019a). Hence, the increasingly available data from low-cost sensor networks has led to new research aimed at combining continuous measurements obtained from a low-cost sensor network with land use data to get spatially and temporally dense air pollution information (Deville Cavellin et al., 2016; Lim et al., 2019; Masiol et al., 2019; Miskell et al., 2018b; Schneider et al., 2017). In Montreal and Vancouver, Canada mobile measurements with low-cost sensors were used to map air pollutants during different seasons (Deville Cavellin et al., 2016) and times of the day (Miskell et al., 2018b), respectively. Schneider et al. (2017) combined air quality data obtained from a low-cost sensor network (24 units) with an urban-scale air quality dispersion model to map $NO_2$ concentrations at near real-time. A network of ten low-cost sensors was used in New York to develop 24 LUR models representative of each hour of the day (Masiol et al., 2019). In a recently published pilot study, we presented another approach to combine $NO_2$ concentrations obtained from a microscale low-cost sensor network (eight sensors) with land use information to identify site and time specific effects of urban design features that disproportionately contribute to population exposure (Weissert et al., 2019a).

Such attempts to fuse land-use and sensor network data face the challenge of demonstrating plausibility of data from low-cost sensor networks (Williams, 2019). A considerable amount of research has focused on sensor performance. A specific issue is drift of sensor signals over time (Clements et al., 2017). Thus, an increasing number of researchers are focusing on developing procedures that allow remote sensor calibrations (Delaine et al., 2019), which is critical for the long-term deployment of large low-cost sensor networks. In our recent work, we developed calibration and remote drift detection procedures for $O_3$ and $NO_2$ sensors deployed in hierarchical networks consisting of well-maintained regulatory sites and low-cost sensors. The approaches were successfully applied and tested at co-located sites (Miskell et al., 2019; Miskell et al., 2016; Miskell et al., 2018a; Weissert et al., 2019a) and were extended to a larger network in Southern California. The corrected sensor data provide a spatially and temporally dense data set of $NO_2$ and $O_3$ concentrations that can claim reliability with an root mean-square error (RMSE) of 7.4 and 5.4 ppb, respectively. Here, first we use this dataset to develop a land-use model for concentrations averaged over two months, and then apply the simple approach described by Weissert et al. (2019a) to model concentrations on an hourly time-scale, both for $NO_2$ and $O_3$. Subsequently, we use an analysis of differences between measured and modelled concentrations to identify local urban conditions that are poorly captured by the static land use model. We show that these deviations have reasonable explanations which in turn reinforces confidence in the original dataset (Williams, 2019).

*Methods*

*Sensor network*

We used the AQY micro air quality monitors from Aeroqual Ltd., Auckland, New Zealand, which are described in detail in Weissert et al. (2019a). The data correction procedures have been comprehensively described elsewhere (Miskell et al., 2019; Miskell et al., 2018a,

Weissert et al., 2019b, c). The low-cost sensor network was deployed in the Inland Empire in Southern California (Fig. 1).

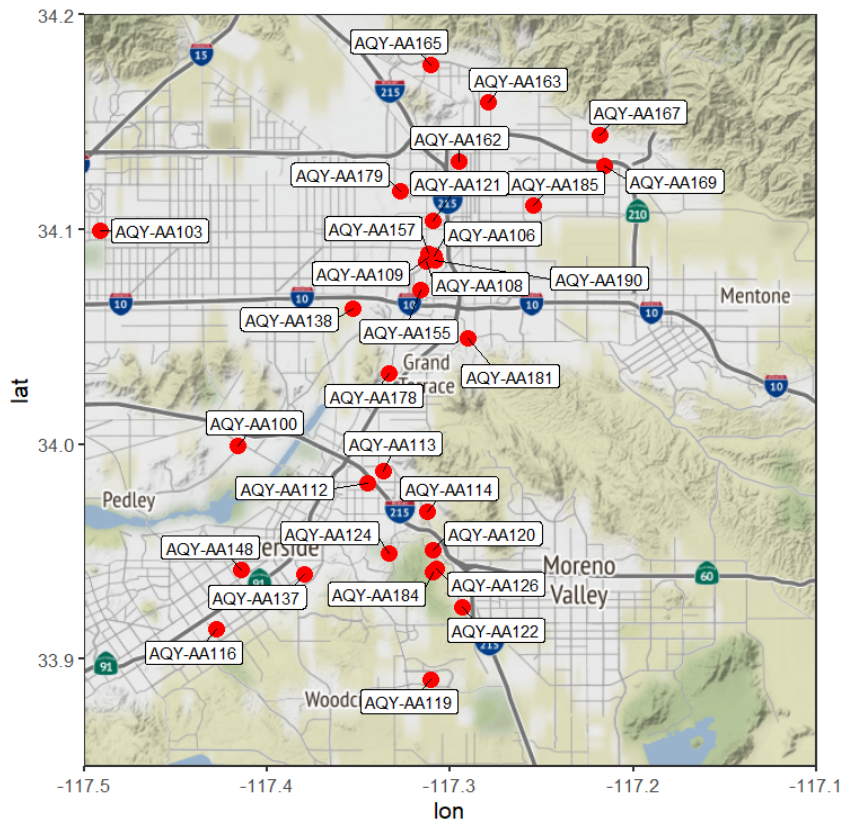

Figure 1. Low-cost sensor sites used for this study ($n = 31$).

For the model building, we used data from 31 low-cost sensor sites during April and May 2018, when most data were available. Pollutant (i.e., $NO_2$ and $O_3$) concentrations were averaged across the two months to develop the 'average' model.

*Predictor variables*

Publicly accessible land use data from Open Street Map and traffic data (Caltrans, 2019) were used as predictor variables. Altitude data was extracted from the SRTM 90m Digital Elevation data (http://srtm.csi.cgiar.org/). We used three variable selection methods to assess the effect of different predictor numbers offered to the model. First, we used all available predictors (Chen et al., 2019), second we optimized buffer distances (Su et al., 2009; Vizcaino and Lavalle, 2018) and third, we removed variables that did not follow the expected direction of effect (Beelen et al., 2013; Vizcaino and Lavalle, 2018). To optimize buffer distances, the correlations between the pollutant concentrations and the predictor variable at each buffer distance were calculated and the one with the highest value of correlation was used in the model.

Table 1. Predictor variables used in the model.

| Predictor variables | Variable code | Unit | Buffer size |
|---|---|---|---|
| Altitude | Elevation | m | |
| Coordinates of the low-cost instrument site | Lat/Long | - | |
| Length of all main roads within the buffer circle | MAJORROADLENGTH | m | 24, 50, 100, 200, 300, 500, 1000 |
| Length of all roads within the buffer circle | ROADLENGTH | m | 24, 50, 100, 200, 300, 500, 1000 |
| Inverse distance to the nearest main road | DISTINVNEAR1 | $m^{-1}$ | |
| Truck traffic | TRUCK_AADT | veh $day^{-1}$ | |
| Vehicle traffic | VEH_AADT | veh $day^{-1}$ | |

*Model building and validation*

We used a random forest (RF) model. RF models have successfully been used in previous studies aiming to predict $NO_2$ concentrations using land use (Araki et al., 2018; Chen et al., 2019; Hu et al., 2017; Zhan et al., 2018). This approach was chosen to minimise the risk of overfitting given that there are relatively few monitoring sites and also to capture non-linear relationships observed between air pollutant concentrations and predictor variables (Araki et al., 2018; Chen et al., 2019; Vizcaino and Lavalle, 2018). RF models are bagged decision tree models, where each tree consists of a random subset of predictor variables from the training dataset and where the final output is the average of multiple decision trees (Breiman, 2001; Grange et al., 2018; Vizcaino and Lavalle, 2018).

We used the caret package in R (v3.5.3) to develop the RF (Kuhn, 2019). Ideally, the data would be split into a training (80%) set, which is used to develop the model, and a test (20%) set, which is used to evaluate the performance of the model. However, given the small sample size of sites, we decided to use all sites to develop the model. Thus, we could not verify the performance of the model on held-out test data. Therefore, the model developed for the monitoring sites may not be representative of other sites. However, the focus of this paper is to assess local effects that result in deviations from the average modelled concentrations, which would not be as affected by the lack of test data. The model is evaluated using a 10-fold cross-validation for resampling where the RMSE is taken as the metric to measure model performance.

*Fusion of the RF model with hourly-averaged data*

To build the model for the temporal variation at the hourly-averaged time-scale, we used the approach described in Weissert et al. (2019a). In brief, we assume that the modelled concentrations ($\bar{C}_{RF,k}$) are linearly related to the hourly-averaged low-cost instrument data ($y_k$) for any given hour on any given day (*l*).

$$y_{k,l} = \hat{a}_{1,l} \bar{C}_{RF,k} + e_{l,k} \qquad (1)$$

where $\hat{a}_{1,l}$ is derived from a least-square regression of eq. 1. An analysis of $e_l$ at the different low-cost sensor sites, *k*, was then used to assess local effects that are not captured by the RF model.

***Results and Discussion***

*Measured pollutant concentrations*

Figure 2 shows boxplots for the $NO_2$ and $O_3$ concentrations measured at the different low-cost sensor sites across the study period which shows that the intra-site variability tends to be larger than the variability between sites for both pollutants. Maximum 8-hour $O_3$ was 120 ppb (site 184). The highest 1-hour average $NO_2$ concentrations were recorded at site 124 (116 ppb).

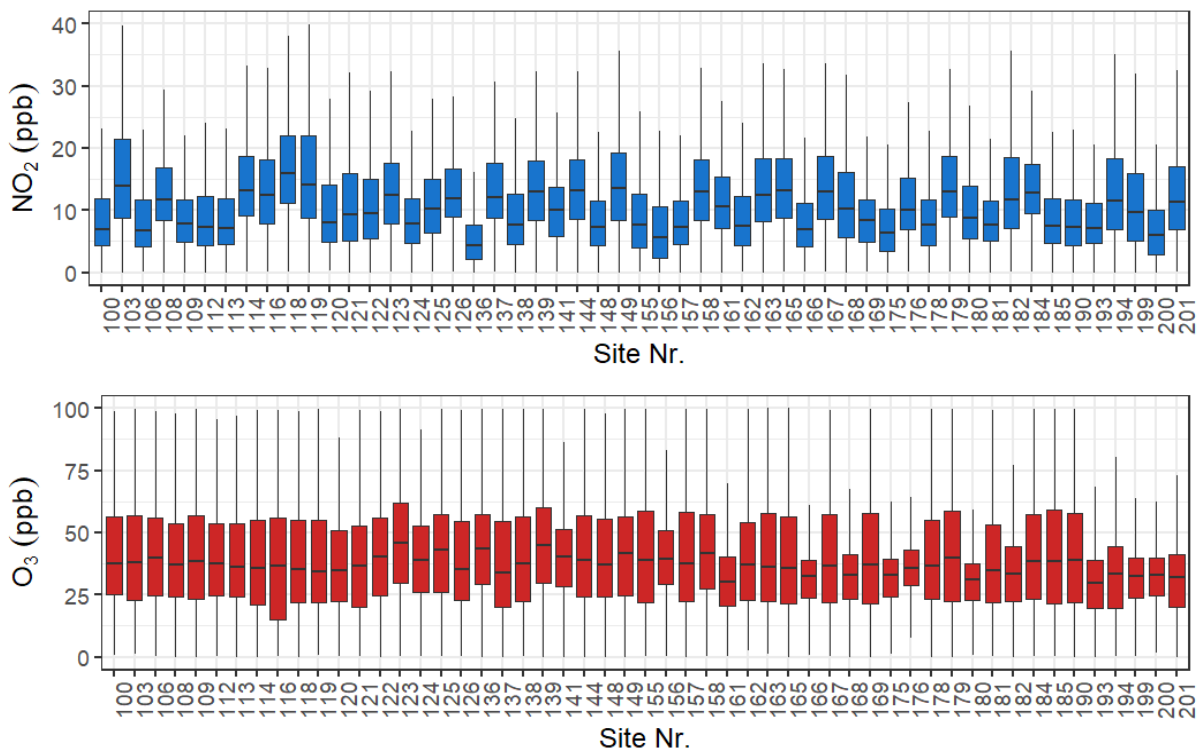

Figure 2. Boxplot for $NO_2$ and $O_3$ concentrations measured at the low-cost sensor sites (x-axis) from April to May 2018. The line denotes the median value. The upper and lower hinges represent the 25th and 75th percentiles. The whiskers extend from the hinge 1.5 times the interquartile range. Outliers are not shown.

Figure 3 shows the spatial variability of mean $O_3$ and $NO_2$ concentrations. Mean $O_3$ concentrations between locations were not highly variable across the region. However, the range of concentrations experienced between locations did differ significantly (figure 2). The mean $NO_2$ concentration was highly spatially variable across the region. The results suggest higher $O_3$ and $NO_2$ concentrations north of Riverside along the mountain range. At other sites, the two pollutants show the expected opposite pattern with higher $NO_2$ concentrations and lower $O_3$ concentration in the south west (SW) direction of Riverside. Although at individual sites the $NO_2$ concentration showed an irregular temporal variation (Weissert et al. 2019b, submitted), on average, with more variability for $NO_2$, the two pollutants showed a simple diurnal variation with the lowest value close to zero: figure 4. For a regular diurnal variation with minimum zero, eq 1 would apply exactly.

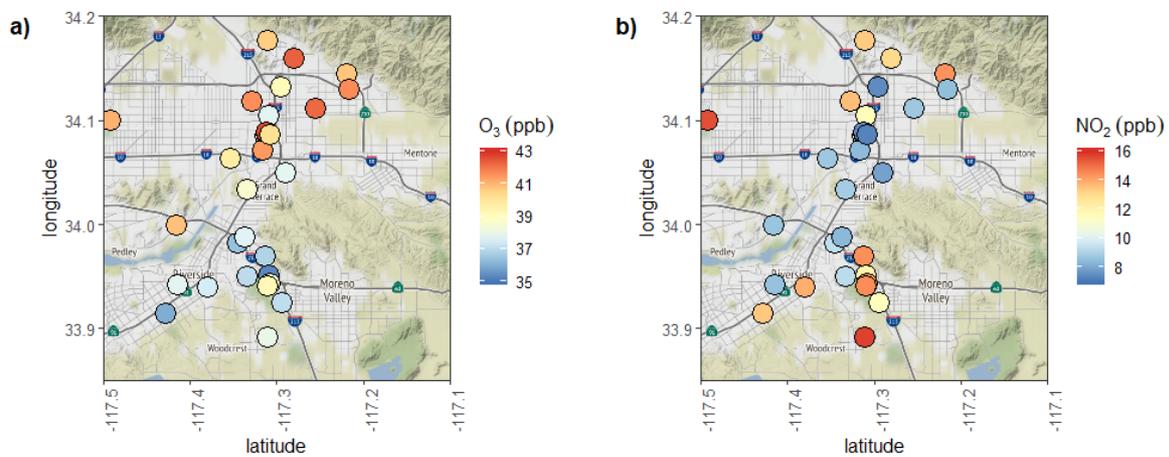

Figure 3. Average measured $O_3$ (left) and $NO_2$ (right) concentrations at the low-cost sensor sites.

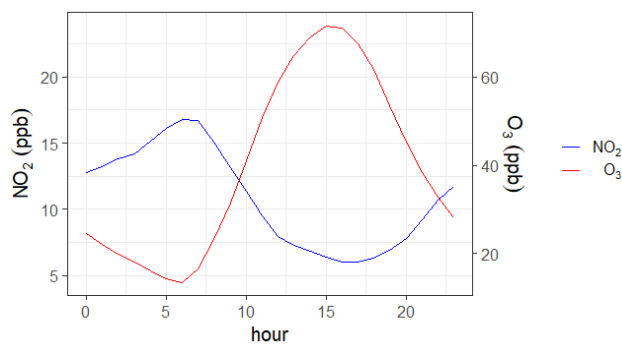

Figure 4. Mean diurnal variation of ozone and nitrogen dioxide, averaged over all sites and days.

*Model results*

Table 2 shows a summary of the resampling results for different predictor selection approaches (1: all predictors, 2: predictors with optimized buffers, 3: predictors with optimized buffers and that follow the expected direction of effect). It shows that the model performed well on the data, with a slightly better performance when using all predictors. When applied to all sensor sites, the $R^2$ between the modelled and measured $NO_2$ and $O_3$ concentrations was 0.93 (RMSE = 1.3 ppb) and 0.73 (RMSE = 1.8 ppb), respectively.

Table 2. Summary of the model performance for different predictor selections (1: all predictors, 2: predictors with optimized buffers, 3: predictors with optimized buffers and that follow the expected direction of effect). $m_{try}$ is the number of variables randomly sampled as candidates.

| Approach | Training data ($n = 31$) | | | | | |
|---|---|---|---|---|---|---|
| | $O_3$ | | | $NO_2$ | | |
| | $R^2$ | RMSE | $m_{try}$ | $R^2$ | RMSE | $m_{try}$ |
| 1 | 0.70 | 1.14 | 11 | 0.71 | 1.82 | 20 |
| 2 | 0.71 | 1.14 | 5 | 0.66 | 1.95 | 2 |
| 3 | NA | NA | NA | 0.66 | 1.95 | 2 |

Figure 5 shows the variable importance derived from the RF suggesting that location (latitude) is the most important predictor for $O_3$, followed by average truck traffic and the inverse distance from the nearest main road. The spatial variability of the measured pollutant concentrations confirms the higher $O_3$ concentrations at higher latitudes at the bottom of the mountain range (Fig. 3). $NO_2$ concentrations were largely dependent on the main road length within 1 km, inverse distance to main road and elevation with a tendency for higher concentrations measured at higher altitudes.

The inverse distance to the nearest main road was also important; however, there was a lot of scatter and the relationship with $NO_2$ concentrations was weak ($R^2 < 0.1$) and following the opposite direction of effect. While predictors following an unexpected direction of effect are excluded in standard linear regression models, RF models may also include predictors with counter-intuitive effects, for example, to compensate for over or under predictions by other predictors (Chen et al., 2019).

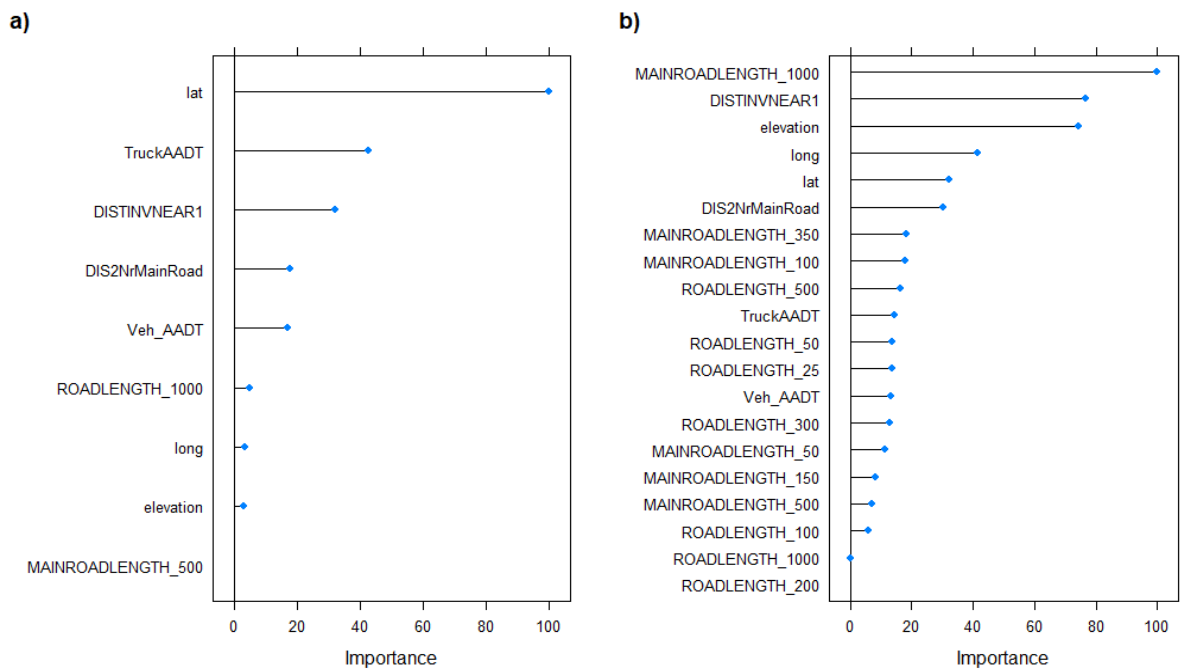

Figure 5. Scaled variable importance (%) plot for the final RF model for a) $O_3$, b) $NO_2$. The variables are listed in order of importance from top to bottom.

*Temporal variation results*

Figure 6 shows the hourly-averaged $O_3$ concentrations measured at the low-cost sensor sites against the modelled and temporally updated $O_3$ concentrations. The figure shows that across the entire study period, the model captured well the temporal variation measured at the low-cost sensor sites. This is to be expected since at most sites and times the variation was a regular diurnal cycle with the lowest value close to zero; hence, the hourly variation would be simply related to the mean.

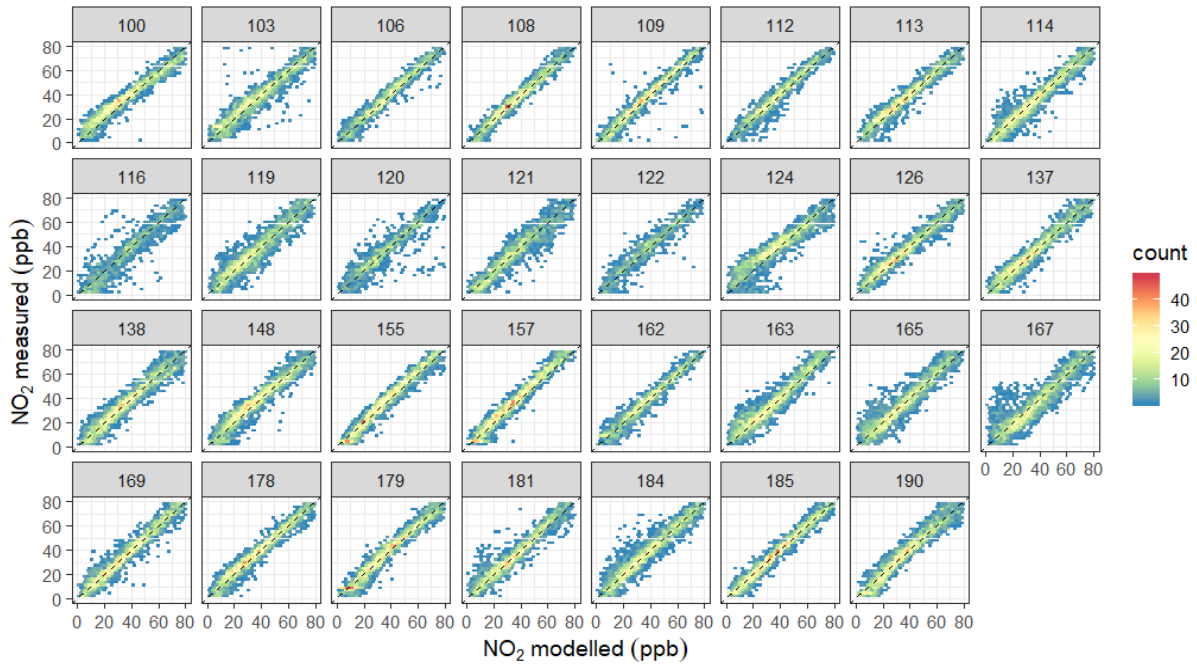

Figure 6. Hexbin plots of hourly averaged $O_3$ concentrations at the low-cost sensor sites against the modelled $O_3$ concentrations (equation 1). The dashed line is the 1:1 line.

Figure 7 shows the same figure for $NO_2$, with the measured $NO_2$ concentrations on the y-axis and the modelled $NO_2$ concentrations on the x-axis. The model captured the overall temporal variability well at most sites, however some deviations can be observed. At site 121, for example, the model was not able to capture the $NO_2$ concentrations measured at this site. Likewise, some high concentrations were missed at site 124.

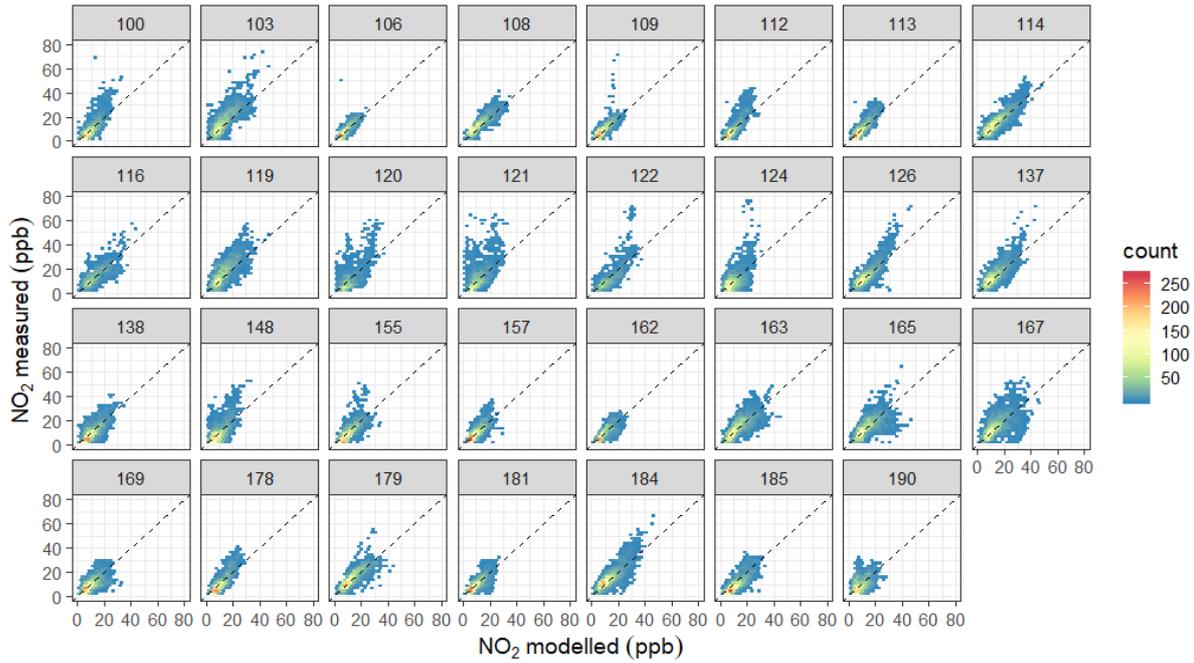

figure 7. Hexbin plots of hourly averaged $NO_2$ concentrations at the low-cost instrument sites against the modelled $NO_2$ concentrations (equation 1). The dashed line is the 1:1 line.

*Analysis of local effects*

Figure 8 shows the unexplained variance for each site as a fraction of the total unexplained variance. $O_3$ concentrations were generally well captured, which is partly due to the lower spatial variability of $O_3$. Sites where the model did not predict temporal $O_3$ concentrations as well include site 120, 124 ,116 and 167. The unexplained variance was slightly higher for $NO_2$ (Fig. 8b) and higher for sites 121, 120, 124 and 167. The spatial variation of the unexplained variance is shown in figure 8c – e suggesting that the model did not perform as well for $O_3$ and $NO_2$ for sites close to the mountain range, north and south of the valley. In an effort to analyse and discuss local effects that may have contributed to unpredicted variations, we also plotted the mean difference term (measured – modelled concentrations) across different wind direction-speed bins (Fig. 9). Some examples for local effects are discussed below.

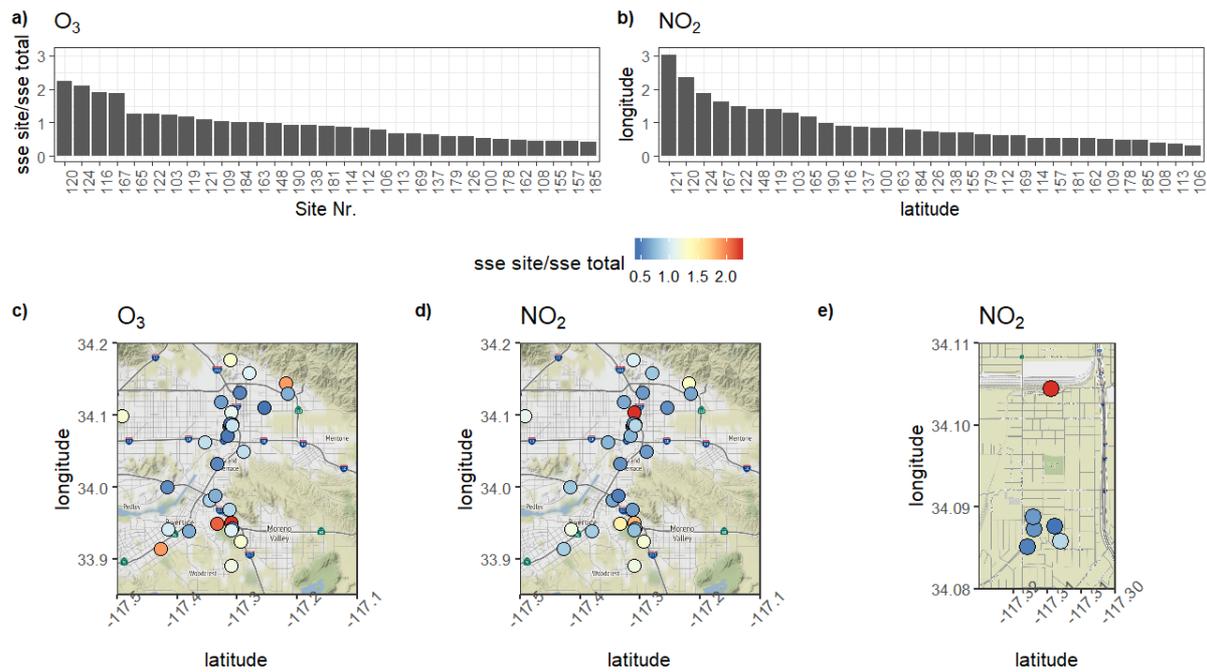

Figure 8. a) – b) $O_3$ and $NO_2$ variation not explained by the model at the different low-cost instrument sites; mean sum of squared error (sse) at the particular site divided by mean sum of squared error (sse total) at all sites. c) – e) maps at different scales to illustrate the spatial variability of the unexplained $NO_2$ variation.

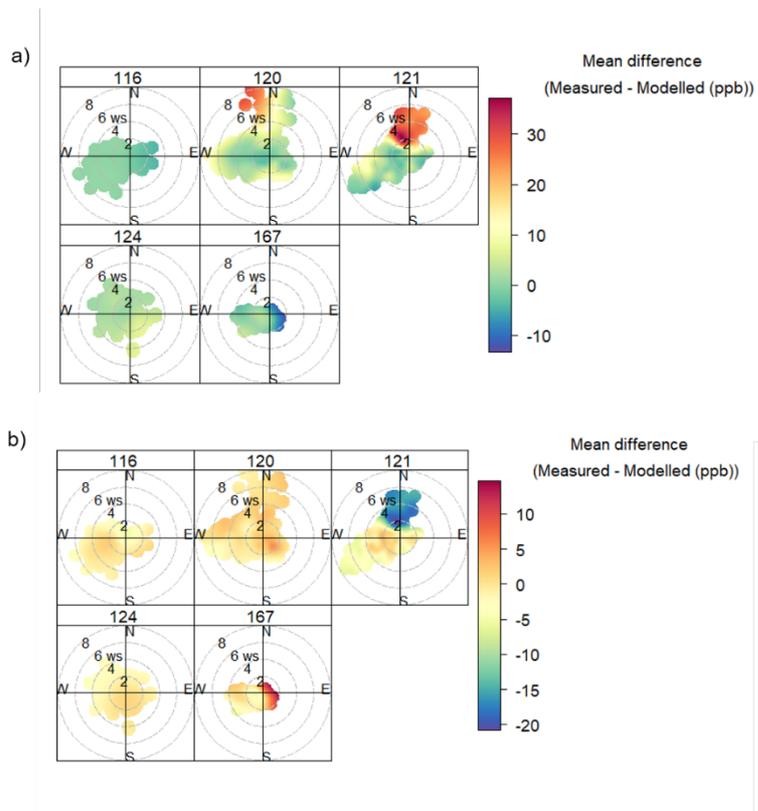

Figure 9. Polar plots showing the mean difference between the measured and modelled $NO_2$ (a) and $O_3$ (b) concentrations divided into different wind direction and wind speed bins.

Site 116 is located SW of Riverside at a high school and the $O_3$ model tended to slightly overestimate $O_3$ concentrations at this site, particularly when wind speed was low. This suggests that $O_3$ concentrations at this school are lower than expected for sites at this latitude and sites with similar traffic and road patterns. Exploration for the reason might focus for example on hyper-local traffic patterns around the site that may be leading to local $O_3$ titration. Site 120 is located in a residential area, east of the Sycamore Canyon Wilderness Park. Thus, the main road length within 1 km, one of the most important predictors for $NO_2$, was relatively small. However, the site was also S/SW of a multi-lane motorway and it is likely that high $NO_2$ and low $O_3$ concentrations were measured when the site was downwind from the motorway as visible in figure 9. Another site that showed distinct differences between measured and modelled $NO_2$ concentrations is site 121. The polar-plot in figure 9 reveals that this was particularly the case when there was north wind direction. This site was located just south of the San Bernardino Santa Fe depot, a major road and rail transport hub (Fig. 8e), which resulted in high $NO_2$ concentrations at this site. Site 124 was located 100 m west from a major road, but measured $O_3$ concentrations were higher and $NO_2$ lower than typically expected within close proximity to a main road. The site is located within a golf-course, suggesting perhaps that nitric oxide (NO) was scavenged by grass and vegetation so $O_3$ titration may have been lower. Site 167 is 400 m south-west from the San Bernardino National Forest, which explains the higher than modelled $O_3$ concentrations and lower than modelled $NO_2$ concentrations associated with wind from the forested area being lower in NO.

To further assess the temporal differences between modelled and measured concentrations, we examined the time-series for April for the sites with disproportionately high unexplained variances (see Figure 10). For $O_3$, the temporally updated model followed the measured $O_3$ concentrations relatively well, although some deviations are observed between the 9th and 14th of April at site 121 where $O_3$ concentrations were lower than modelled. This site also showed large deviations for $NO_2$ for the same time period, suggesting the local emission sources may have changed during this time period. Wind data for this period showed a change in wind direction from dominating south-westerlies to northerlies. Thus, the influence from the train depot north of this site would therefore have been stronger between the 9th and 14th of April, particularly on the 9th and 13th when measured $NO_2$ concentrations were considerably higher than modelled. Similarly, site 167 showed larger deviations between the 9th and 14th of April, when measured air was mostly coming from the San Bernardino National Forest.

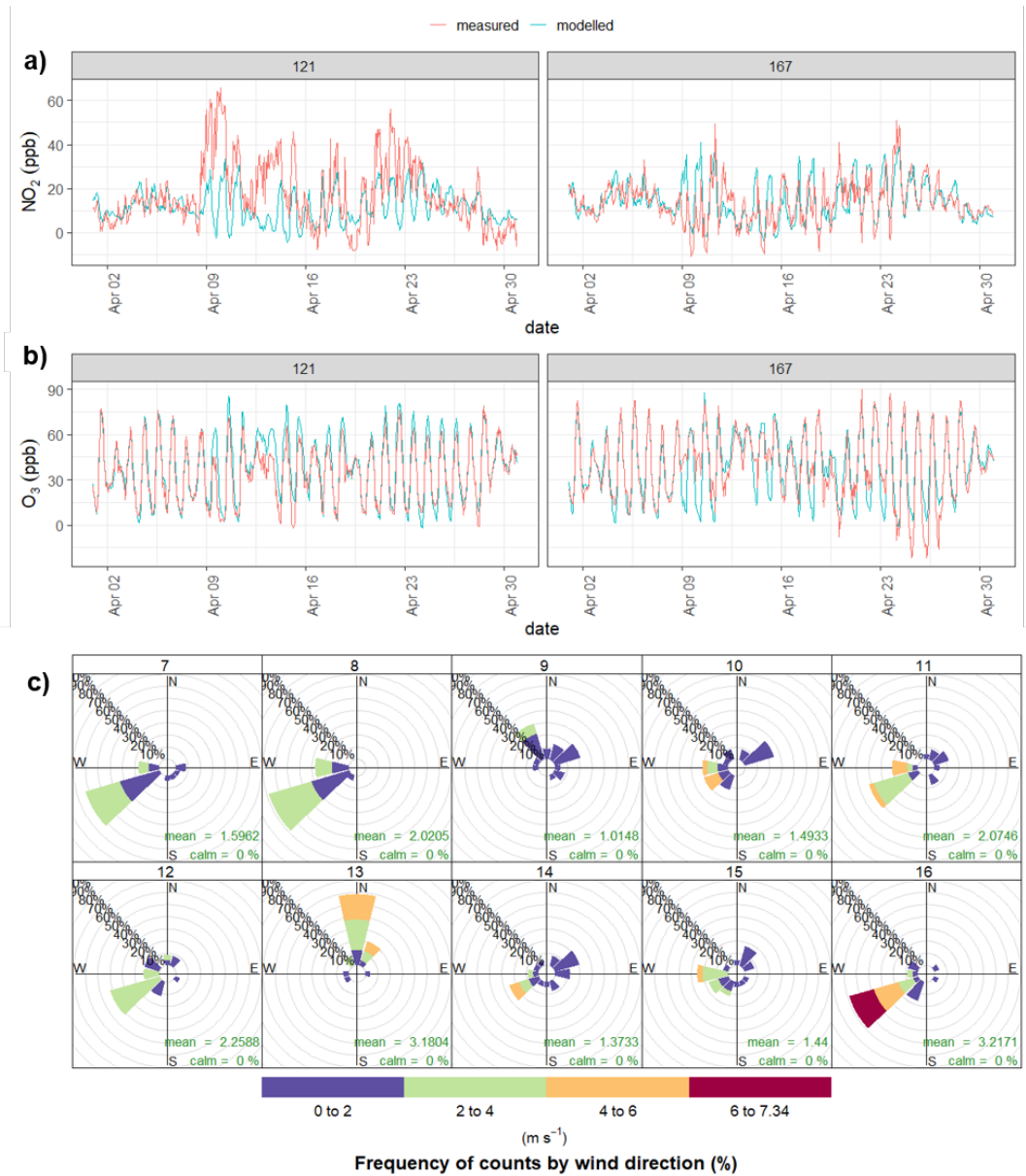

Figure 10. a) hourly measured NO$_2$ concentrations and b) hourly measured O$_3$ concentrations during April, c) frequency of wind speed and wind direction in different wind speed and wind direction categories at site 121 between the 7th and 16th of April (the panels are different days).

*Temporally variable pollution maps*

Finally, the presented approach allows mapping pollutant concentrations for any given day and hour measurement data are available. An example for predicted NO$_2$ concentrations for 10/05/2018 at 07:00 local time is shown in figure 11. Given that the prediction success of the model developed here is limited due to the relatively small number of sites, the figure is focused on an area where measurements were relatively dense and the predictions likely more representative. It shows the expected high NO$_2$ concentrations during the morning rush hour. Spatially, NO$_2$ concentrations were slightly higher north and south of the study area, which is partially due to higher altitudes. However, overall spatial differences were small. The spatial coverage and prediction power of the model may be improved by supplementing low-cost sensor networks with diffusion tube campaigns involving local communities or schools. This could then allow mapping pollutant concentrations with more confidence for any given day and hour at a neighbourhood scale and offer insights about pollution hotspots and their temporal variation.

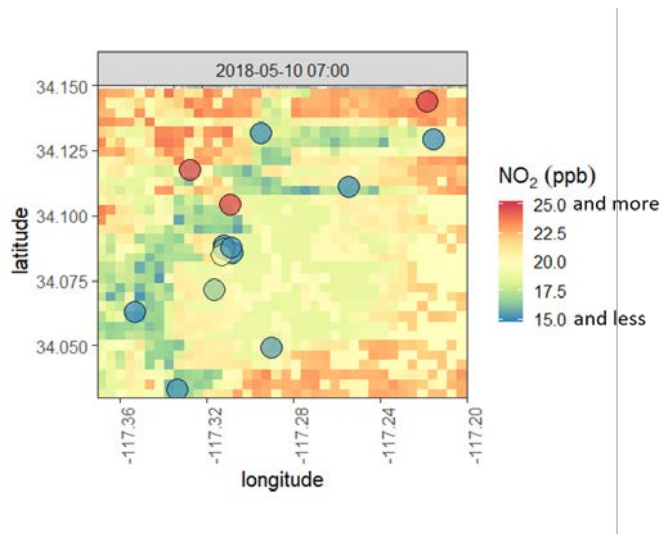

Figure 11. Predicted NO$_2$ concentrations at 500 m resolution for northern Riverside, where the low-cost sensor network was the densest network in the region, with measured NO$_2$ concentrations at the sensor sites superimposed: 10$^{th}$ May 2018 at 07:00.

*Conclusions*

This paper has presented results of a RF model to predict concentration values for $O_3$ and $NO_2$ based on LUR and a low-cost sensor network that was deployed in the Inland Empire region in Southern California. A previously described procedure was used to remotely calibrate the low-cost sensors using data from the more sparsely distributed regulatory network. We combined land use information and an RF model with hourly low-cost sensor data to identify local effects on $O_3$ and $NO_2$ concentrations at a high temporal resolution. The mean RF models performed well for $NO_2$ and $O_3$ concentrations ($R^2$ = 0.93/RMSE = 1.3 ppb and $R^2$ = 0.73/RMSE = 1.8 ppb, respectively) at the low-cost sensor sites. The mean modelled pollutant concentrations were successfully updated hourly using the low-cost instrument data. The model for $O_3$ combined with the low-cost instrument data captured the spatial and temporal variation well. For $NO_2$, variations from the model highlighted particular urban features, not accounted for by the general land-use modelling, that under particular circumstances resulted in significantly increased pollutant concentration. The model proved to be an effective and simple way to highlight the spatial and temporal distribution of local effects that may disproportionately contribute to pollutant concentrations. The findings may be associated with different meteorological conditions (e.g. higher pollutant concentrations expected for particular wind directions) offering support for local pollution alerts. If supplemented with more dense measurements, for example using diffusion tube campaigns, the model would allow for mapping pollutant concentrations for any given day and hour, which may be updated in near real-time as long as measurement data are available.

**Acknowledgement**

This work was funded by the New Zealand Ministry for Business, Innovation and Employment, contract UOAX1413. This work was performed in collaboration with the Air Quality Sensor Performance Evaluation Center (AQ-SPEC) at the South Coast Air Quality Management District (South Coast AQMD). The authors would like to acknowledge the work of Mr. Berj

Der Boghossian for his technical assistance with deploying AQY sensor nodes. The authors would like to acknowledge the work of the South Coast AQMD Atmospheric Measurements group of dedicated instrument specialists that operate, maintain, calibrate, and repair air monitoring instrumentation to produce regulatory-grade air monitoring data. DEW acknowledges the support of a fellowship program at the Institute of Advanced Studies, Durham University, UK.

## 6. Competing interests

LW, EM, KA and GSH are employees of Aeroqual Ltd, manufacturer of the sensor nodes used in the study. GSH and DEW are founders and shareholders in Aeroqual Ltd.

*References*

Araki, S., Shima, M., Yamamoto, K., 2018. Spatiotemporal land use random forest model for estimating metropolitan $NO_2$ exposure in Japan. Sci Total Environ 634, 1269-1277.

Beelen, R., Hoek, G., Vienneau, D., Eeftens, M., Dimakopoulou, K., Pedeli, X., Tsai, M.-Y., Künzli, N., Schikowski, T., Marcon, A., Eriksen, K.T., Raaschou-Nielsen, O., Stephanou, E., Patelarou, E., Lanki, T., Yli-Tuomi, T., Declercq, C., Falq, G., Stempfelet, M., Birk, M., Cyrys, J., von Klot, S., Nádor, G., Varró, M.J., Dėdelė, A., Gražulevičienė, R., Mölter, A., Lindley, S., Madsen, C., Cesaroni, G., Ranzi, A., Badaloni, C., Hoffmann, B., Nonnemacher, M., Krämer, U., Kuhlbusch, T., Cirach, M., de Nazelle, A., Nieuwenhuijsen, M., Bellander, T., Korek, M., Olsson, D., Strömgren, M., Dons, E., Jerrett, M., Fischer, P., Wang, M., Brunekreef, B., de Hoogh, K., 2013. Development of $NO_2$ and $NO_x$ land use regression models for estimating air pollution exposure in 36 study areas in Europe – The ESCAPE project. Atmospheric Environment 72, 10-23.

Breiman, L., 2001. Random Forests. Machine Learning 45, 5 - 32.

Brokamp, C., Jandarov, R., Rao, M.B., LeMasters, G., Ryan, P., 2017. Exposure assessment models for elemental components of particulate matter in an urban environment: A comparison of regression and random forest approaches. Atmos Environ (1994) 151, 1-11.

Caltrans (2019). Caltrans GIS Data – Truck Volumes AADT. https://gisdata-caltrans.opendata.arcgis.com/datasets/dfe7fd95282946db98145e9bcaf710fb_0, Accessed: September, 2019).

Chen, J., de Hoogh, K., Gulliver, J., Hoffmann, B., Hertel, O., Ketzel, M., Bauwelinck, M., van Donkelaar, A., Hvidtfeldt, U.A., Katsouyanni, K., Janssen, N.A.H., Martin, R.V., Samoli, E., Schwartz, P.E., Stafoggia, M., Bellander, T., Strak, M., Wolf, K., Vienneau, D., Vermeulen, R., Brunekreef, B., Hoek, G., 2019. A comparison of linear regression, regularization, and machine learning algorithms to develop Europe-wide spatial models of fine particles and nitrogen dioxide. Environ Int 130, 104934.

Clements, A.L., Griswold, W.G., Rs, A., Johnston, J.E., Herting, M.M., Thorson, J., Collier-Oxandale, A., Hannigan, M., 2017. Low-Cost Air Quality Monitoring Tools: From Research to Practice (A Workshop Summary). Sensors (Basel) 17.

Delaine, F., Lebental, B., Rivano, H., 2019. In Situ Calibration Algorithms for Environmental Sensor Networks: A Review. IEEE Sensors Journal 19, 5968-5978.

Deville Cavellin, L., Weichenthal, S., Tack, R., Ragettli, M.S., Smargiassi, A., Hatzopoulou, M., 2016. Investigating the Use Of Portable Air Pollution Sensors to Capture the Spatial Variability Of Traffic-Related Air Pollution. Environ Sci Technol 50, 313-320.

Epstein, S.A., Lee, S., Katzenstein, A.S., Carreras-Sospedra, M., Zhang, X., Farina, S.C., Vahmani, P., Fine, P.M., Ban-Weiss, G., 2017. Air-quality implications of widespread adoption of cool roofs on ozone and particulate matter in southern California. PNAS 114, 8991-8996.

Feinberg, S.N., Williams, R., Hagler, G., Low, J., Smith, L., Brown, R., Garver, D., Davis, M., Morton, M., Schaefer, J., Campbell, J., 2019. Examining spatiotemporal variability of urban particulate matter and application of high-time resolution data from a network of low-cost air pollution sensors. Atmospheric Environment 213, 579-584.

Grange, S.K., Carslaw, D.C., Lewis, A.C., Boleti, E., Hueglin, C., 2018. Random forest meteorological normalisation models for Swiss $PM_{10}$ trend analysis. Atmospheric Chemistry and Physics 18, 6223-6239.

Hoek, G., Beelen, R., de Hoogh, K., Vienneau, D., Gulliver, J., Fischer, P., Briggs, D., 2008. A review of land-use regresssion models to assess spatial variation of outdoor air pollution Atmospheric Environment 42, 7561 - 7578.

Hu, X., Belle, J.H., Meng, X., Wildani, A., Waller, L.A., Strickland, M.J., Liu, Y., 2017. Estimating PM2.5 Concentrations in the Conterminous United States Using the Random Forest Approach. Environ Sci Technol 51, 6936-6944.

Kuhn, M., 2019. caret: Classification and Regression Training, R package version 6.0-84.

Kumar, A., Singh, D., Singh, B.P., Singh, M., Anandam, K., Kumar, K., Jain, V.K., 2015. Spatial and temporal variability of surface ozone and nitrogen oxides in urban and rural ambient air of Delhi-NCR, India. Air Quality, Atmosphere & Health 8, 391-399.

Li, H.Z., Gu, P., Ye, Q., Zimmerman, N., Robinson, E.S., Subramanian, R., Apte, J.S., Robinson, A.L., Presto, A.A., 2019a. Spatially dense air pollutant sampling:

Implications of spatial variability on the representativeness of stationary air pollutant monitors. Atmospheric Environment: X 2, 100012.

Li, L., Girguis, M., Lurmann, F., Wu, J., Urman, R., Rappaport, E., Ritz, B., Franklin, M., Breton, C., Gilliland, F., Habre, R., 2019b. Cluster-based bagging of constrained mixed-effects models for high spatiotemporal resolution nitrogen oxides prediction over large regions. Environ Int 128, 310-323.

Lim, C.C., Kim, H., Vilcassim, M.J.R., Thurston, G.D., Gordon, T., Chen, L.C., Lee, K., Heimbinder, M., Kim, S.Y., 2019. Mapping urban air quality using mobile sampling with low-cost sensors and machine learning in Seoul, South Korea. Environ Int 131, 105022.

Masiol, M., Squizzato, S., Chalupa, D., Rich, D.Q., Hopke, P.K., 2019. Spatial-temporal variations of summertime ozone concentrations across a metropolitan area using a network of low-cost monitors to develop 24 hourly land-use regression models. Sci Total Environ 654, 1167-1178.

Masiol, M., Zikova, N., Chalupa, D.C., Rich, D.Q., Ferro, A.R., Hopke, P.K., 2018. Hourly land-use regression models based on low-cost PM monitor data. Environ Res 167, 7-14.

Miskell, G., Alberti, K., Feenstra, B., Henshaw, G., Papapostolou, V., Patel, H., Polidori, A., Salmond, J.A., Weissert, L.F., Williams, D.E., 2019. Reliable data from low-cost ozone sensors in a hierarchical network, http://arxiv.org/abs/1906.08421.

Miskell, G., Salmond, J., Alavi-Shoshtari, M., Bart, M., Ainslie, B., Grange, S., McKendry, I.G., Henshaw, G.S., Williams, D.E., 2016. Data Verification Tools for Minimizing Management Costs of Dense Air-Quality Monitoring Networks. Environ Sci Technol 50, 835-846.

Miskell, G., Salmond, J.A., Williams, D.E., 2018a. Solution to the Problem of Calibration of Low-Cost Air Quality Measurement Sensors in Networks. ACS Sensors 3, 832-843.

Miskell, G., Salmond, J.A., Williams, D.E., 2018b. Use of a handheld low-cost sensor to explore the effect of urban design features on local-scale spatial and temporal air quality variability. Science of The Total Environment 619-620, 480-490.

Popoola, O.A.M., Carruthers, D., Lad, C., Bright, V.B., Mead, M.I., Stettler, M.E.J., Saffell, J.R., Jones, R.L., 2018. Use of networks of low cost air quality sensors to quantify air quality in urban settings. Atmospheric Environment 194, 58-70.

Schneider, P., Castell, N., Vogt, M., Dauge, F.R., Lahoz, W.A., Bartonova, A., 2017. Mapping urban air quality in near real-time using observations from low-cost sensors and model information. Environment International 106, 234-247.

Son, Y., Osornio-Vargas, Á.R., O'Neill, M.S., Hystad, P., Texcalac-Sangrador, J.L., Ohman-Strickland, P., Meng, Q., Schwander, S., 2018. Land use regression models to assess air pollution exposure in Mexico City using finer spatial and temporal input parameters. Science of The Total Environment 639, 40-48.

South Coast AQMD, 2016. Final 2016 - Air Quality Management Plan.

Su, J.G., Jerrett, M., Beckerman, B., 2009. A distance-decay variable selection strategy for land use regression modeling of ambient air pollution exposures. Sci Total Environ 407, 3890-3898.

Vizcaino, P., Lavalle, C., 2018. Development of European NO2 Land Use Regression Model for present and future exposure assessment: Implications for policy analysis. Environ Pollut 240, 140-154.

Weissert, L.F., Alberti, K., Miskell, G., Pattinson, W., Salmond, J.A., Henshaw, G., Williams, D.E., 2019a. Low-cost sensors and microscale land use regression: Data fusion to resolve air quality variations with high spatial and temporal resolution. Atmospheric Environment 213, 285-295.

Weissert, L.F., Miskell, G., Miles, E., Feenstra, B., Papapostolou, V., Polidori, A., Henshaw, G.S., Salmond, J.A., Williams, D.E., 2019b. Hierarchical network design for nitrogen dioxide measurement in urban environments, part 1: proxy selection. Submitted to Atmospheric Environment.

Weissert, L.F., Miles, E., Miskell, G., Alberti, K., Feenstra, B., Henshaw, G.S., Papapostolou, V., Patel, H., Polidori, A., Polidori, A., Salmond, J.A., Williams, D.E, 2019c. Hierarchical network design for nitrogen dioxide measurement in urban environments, part 2: network-based sensor calibration. Submitted to Atmospheric Environment.

Williams, D.E., 2019. Low cost sensor networks: How do we know the data are reliable? ACS Sensors ; https://doi.org/10.1021/acssensors.9b01455.

Yeganeh, B., Hewson, M.G., Clifford, S., Tavassoli, A., Knibbs, L.D., Morawska, L., 2018. Estimating the spatiotemporal variation of NO2 concentration using an adaptive neuro-fuzzy inference system. Environmental Modelling & Software 100, 222-235.

Zhan, Y., Luo, Y., Deng, X., Zhang, K., Zhang, M., Grieneisen, M.L., Di, B., 2018. Satellite-Based Estimates of Daily NO2 Exposure in China Using Hybrid Random Forest and Spatio-temporal Kriging Model. Environ Sci Technol 52, 4180-4189.